\documentclass{Interspeech2024}

\usepackage{comment}
\usepackage{siunitx}
\usepackage{multirow}

\interspeechcameraready

\title{Spatial Voice Conversion: Voice Conversion Preserving \\ Spatial Information and Non-target Signals}

\name[affiliation={1}]{Kentaro}{Seki}
\name[affiliation={1,2}]{Shinnosuke}{Takamichi}
\name[affiliation={1}]{Norihiro}{Takamune}
\name[affiliation={1}]{Yuki}{Saito}
\name[affiliation={1}]{Kanami}{Imamura}
\name[affiliation={1}]{Hiroshi}{Saruwatari}

\address{
  $^1$The University of Tokyo, Japan \,\,\,
  $^2$Keio University, Japan}
\email{seki-kentaro922@g.ecc.u-tokyo.ac.jp}

\keywords{voice conversion, speech extraction, blind source separation, augmented reality, virtual reality}

\newcommand{\mean}[1]{\mathbb{E}\left[#1\right]}

\newcommand{\ideal}{Ideal}
\newcommand{\steering}{Steering}
\newcommand{\inverse}{Inverse}

\begin{document}

\maketitle
\setlength{\abovedisplayskip}{3pt} % 式の上部のマージン
\setlength{\belowdisplayskip}{3pt} % 式の下部のマージン

\begin{abstract}
This paper proposes a new task called \textit{spatial voice conversion}, which aims to convert a target voice while preserving spatial information and non-target signals. 
Traditional voice conversion methods focus on single-channel waveforms, ignoring the stereo listening experience inherent in human hearing. 
Our baseline approach addresses this gap by integrating blind source separation (BSS), voice conversion (VC), and spatial mixing to handle multi-channel waveforms. 
Through experimental evaluations, we organize and identify the key challenges inherent in this task, such as maintaining audio quality and accurately preserving spatial information. 
Our results highlight the fundamental difficulties in balancing these aspects, providing a benchmark for future research in spatial voice conversion. 
The proposed method's code is publicly available to encourage further exploration in this domain.
\end{abstract}

\section{Introduction}
We humans possess the ability for auditory attention in speech communication, known as the \textit{cocktail party effect}~\cite{cherry1953some}.
One contributing factor to this phenomenon is \textit{spatial attention}~\cite{blauert1997spatial,ebata2003spatial}, the capability to focus on the voice coming from the desired direction among sounds heard with both ears.
Another contributing factor is \textit{informative attention}.
For instance, if the voice is attractive to the listeners, we tend to listen more attentively~\cite{amitava03hearingvoice}. 
Additionally, if the voice has an unfamiliar pronunciation, we interpret it as a voice with an understandable pronunciation to comprehend the spoken content~\cite{yevgeniy21adaptation-to-english}. 
It is important to note that \textit{the ability is not about selection but attention}. 
In other words, for perceiving and comprehending the target speech, our ability involves processing the speech of unattended speakers as well~\cite{lewis1970semantic,bentin1995semantic}.
For example, even when paying attention to the pronunciation information of a speaker, one still understands the spatial location of that speaker and the information of other speakers without altering their spatial context. 

Then, \textit{is it possible to emulate this human ability with computers?} 
That is, we aim to develop a speech processing technology that transforms only the desired voice information while retaining all other content.
If this can be achieved, it would enable communication augmentation that emphasizes human auditory attention more in the real world or in virtual reality (VR) spaces. 

\begin{figure}[t]
    \centering
    \includegraphics[width=0.7\linewidth]{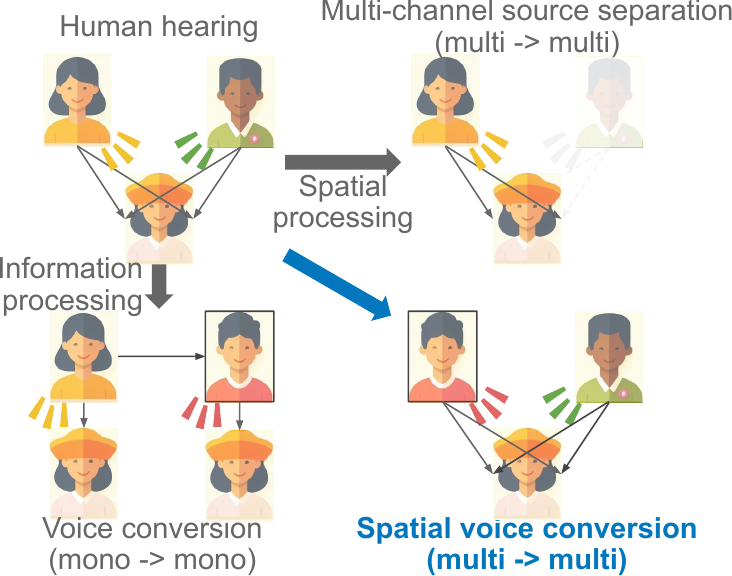}
    \vspace{-3mm}
    \caption{Spatial voice conversion. It converts target-speaker signal in multi-channel observations while unchanging spatial information and non-target signals.}
    \label{fig:overview}
    \vspace{-5mm}
\end{figure}

% これに関連する技術がマルチチャネル音源分離~\cite{ref}と音声変換である．マルチチャネル音源分離は，複数音源を空間的に分離する技術であり，例えば音声認識の front-end として音源選択に利用される~\cite{ref}．対して音声変換は，発話内容を保持したまま対象音声を別の音声に変換する技術であり，例えば外国語発音の変換や刺激の強い感情の抑制などに使用される~\cite{ref}．この2つの技術を結合することで，対象音声のみを変換することはできる．すなわち，対象音声を空間的に分離して抽出したのち，その抽出音声を変換する．しかしながら，前述の目標を達成するにあたりこれらの技術だけでは不十分である．何故ならば，対象音声以外の情報は保持されなければならず，また，全体の性能は，変換音声のみならず，変換音声のみが変更された混合音声について評価されるべきである．
Related technologies include multi-channel audio source separation~\cite{makino2018audio} and voice conversion~\cite{sisman2020overview}. Multi-channel audio source separation (upper right of Fig.~\ref{fig:overview}) is a technology that spatially separates multiple audio sources, for example, used as source selection in the front-end of automatic speech recognition~\cite{sainath2017multichannel,masuyama2023exploring}. On the other hand, voice conversion (lower left of Fig.~\ref{fig:overview}) is a technology that converts targeted speech to another speech while retaining the spoken content, used for purposes such as converting foreign accents~\cite{aryal14vc-nonnative-accent} or neutralize speech emotion~\cite{zhou2022emotional}. By combining these two technologies, it is possible to convert only the targeted speech. That is, the targeted speech is spatially separated and extracted, then the extracted speech is converted. However, these technologies alone are insufficient to achieve the aforementioned goal. Because information other than the target speech must be retained, and the overall performance should be evaluated not only for the converted speech but also for the mixed speech with only the converted speech changed.

% そこで本稿では，target speaker voice conversion という新たなタスクを定義し，その baseline 構築および評価方法について議論する．このタスクでは，マルチチャネル音声のうち対象話者の音声だけを変換し，それ以外の内容（例えば，非対象話者，空間位置）を保持する．すなわち，入力はマルチチャネル音声，出力は対象話者音声だけを変換したマルチチャネル音声であり，評価対象はその出力マルチチャネル音声である．Baseline は，spatial demixing, target speaker identification, voice conversion, spatial mixing のモジュールから成る．評価方法については，対象話者音声だけを変換したマルチチャネル音声を評価する軸を検討するとともに，各モジュールのエラーが propagate した際の知覚的な影響を調査する．これらを通して，本タスクにおいて解決すべき問題と目指すべき指針を策定する．
% Regarding the evaluation methods, we will consider the axis for evaluating the multi-channel audio with only the target speaker's voice converted, as well as investigate the perceptual impact when errors from each module propagate. Through these, we will formulate the problems to be solved and the guidelines to aim for in this task.
% 複数のモジュールを結合することで生じる問題をリストし、本タスクが単なるblind source separationとvoice conversionの結合よりも難しいタスクであることを指摘する。その課題を解決する方法を一つ提案し、実験的評価において本タスクのchallengingな課題を明らかにする。
Therefore, this paper defines a new task called \textit{spatial voice conversion} (lower right of Fig.~\ref{fig:overview}) and organizes the challenges of this task through experiments using our proposed baseline method. 
In this task, only the speech of the target speaker in the multi-channel audio is converted, while retaining all other content (e.g., non-target speakers, spatial location). 
That is, the input is multi-channel audio, and the output is multi-channel audio with only the target speaker's voice converted. 
The subject of evaluation is this multi-channel output audio. 
Our proposed baseline method consists of modules for blind source separation, voice conversion, and spatial mixing. 
Experiments demonstrate that this task involves fundamentally more challenging issues than simply combining modules. 
We publish the code of our method\footnote{\url{https://github.com/sarulab-speech/spatial_voice_conversion.git}} as a baseline for future research on this task.

\section{Related work}

\label{sec:related_work}
\textbf{Multi-channel source separation.} 
\begin{comment}    
Array signal processing is a method that computationally implements spatial attention using microphone arrays. 
Unlike delay-and-sum beamformers~\cite{van1988beamforming}, blind source separation (BSS) does not require assumptions about spatial information. 
The BSS methods assume statistical independence among different signal sources to construct separation filters~\cite{kim2006independent,hiroe2006solution,ono2011stable}.
Furthermore, methods combining prior information about spatial details with BSS have been proposed~\cite{parra2002geometric,knaak2007geometrically}, such as geometrically constrained independent vector analysis~(GC-IVA)~\cite{li2020geometrically}. BSS can also be used for speech enhancement. That is, it involves extracting only the targeted speech from multi-channel sources.%~\cite{具体例が欲しい}.
%%%%%%%%%%%%%
Array signal processing is a method that computationally implements spatial attention using microphone arrays. 
Unlike delay-and-sum beamformers~\cite{van1988beamforming}, 信号源の統計的独立性に基づいてassumptions about spatial informationをrequireすることなくspatial attentionを実現する方法がblind source separation(BSS)の一種として提案されている~\cite{kim2006independent,hiroe2006solution,ono2011stable}．
さらに，spatial informationに関する仮定であって誤差を含む仮定をBSSと統合する方法が提案されている．
\end{comment}
Array signal processing is a method that computationally implements spatial attention using microphone arrays. 
Unlike delay-and-sum beamformers~\cite{van1988beamforming}, methods for achieving spatial attention without requiring assumptions about the statistical independence of signal sources have been proposed under the framework of blind source separation (BSS)~\cite{kim2006independent,hiroe2006solution,ono2011stable}. 
Additionally, methods have been proposed that combine BSS with prior spatial information, which may contain errors~\cite{parra2002geometric,knaak2007geometrically}, such as geometrically constrained independent vector analysis (GC-IVA)~\cite{li2020geometrically}. 
Additionally, BSS can be used for speech enhancement by extracting only the targeted speech from multi-channel sources.

%% 情報注意：シングルチャネル音声強調/分離
% 情報注意を計算機的に実装する方法として，シングルチャネル信号に対する音声強調・音源分離手法が研究されている．近年は機械学習を用いる方法が高い性能を達成しており，Conv-TAS NETやdual path RNNなどのモデルが提案されている．
\begin{comment}
For the computational implementation of information attention, research has been conducted on speech enhancement and source separation methods for single-channel signals.
Certainly! Here's the translation:
In recent years, methods utilizing machine learning have achieved high performance, with models such as Conv-TAS NET~\cite{luo2019conv} and dual path RNN~\cite{luo2020dual} being proposed.
\end{comment}
%% VC：タスクを説明し，色々なSoTAを紹介
% Voice Conversion は，音声のパラ言語情報・非言語情報を変換するタスクである while preserving 言語情報. 近年は機械学習モデルを用いた方法が盛んに研究されており，GANを用いる方法やdiffusionを用いる方法などが提案されている． 
\textbf{Voice conversion.} Voice conversion (VC) is a task that involves transforming the paralinguistic and non-linguistic information of speech while preserving the linguistic content. Various approaches based on deep learning have been proposed, including diffusion models~\cite{liu2021diffsvc, popov2022diffusionbased}. Essentially, the VC task assumes that the input is a monaural signal containing only the targeted voice. If non-target audio (e.g., background noise) is included, it is often removed so that only the targeted voice is converted~\cite{valentini2016investigating}.

% 本研究と同様に，音声情報以外の情報を保存する音声処理の研究が行われている．voice conversion において背景雑音を保つタスクとしてnoisy-to-noisy voice conversionが提案されている．また，音声強調において空間情報を保つ方法が提案されている．
\begin{comment}
Like this study, research has explored audio processing tasks that involve dealing information beyond just the speech content. For instance, in the domain of voice conversion, there is the task of noisy-to-noisy voice conversion~\cite{xie2021noisy}, where the goal is to maintain background noise during the conversion process. This is a notable extension beyond traditional voice conversion, as it considers and retains non-speech information, enhancing the realism of the converted speech.
Similarly, in the context of speech enhancement, methods have been proposed to preserve spatial information~\cite{togami2024real}.
These approaches aim to enhance the speech signal while maintaining the spatial characteristics of the audio, providing a more comprehensive and context-aware improvement. The convergence of these research areas demonstrates a growing interest in audio processing techniques that go beyond speech content alone and consider additional environmental factors for more realistic and effective processing.
\end{comment}
\textbf{Preserving non-target information.} Unlike BSS-based speech enhancement and VC, which remove non-target information, several methods have been proposed to reflect non-target information in the processed speech. In noisy-to-noisy VC~\cite{xie2021noisy}, the background noise contained in the input speech is reflected in the converted speech. Another paper on speech enhancement~\cite{togami2024real} outputs stereo speech that not only emphasizes the target speech contained in stereo audio but also preserves its spatial information. These studies value the context in which the targeted speech occurs, taking the stance that information other than the targeted speech should be preserved, not removed. Our research also adopts this stance. The convergence of this direction demonstrates a growing interest in audio processing techniques that go beyond targeted speech alone.
% Non-target 情報を除去する BSS-based speech enhancement や VC と異なり，non-target 情報を処理後音声に反映する方法が幾つか提案されている．例えば，noisy-to-noisy VC では，モノラルの入力音声に含まれる背景雑音を，変換後音声に反映する．音声強調における別の論文では，ステレオ音声に含まれる target speech を強調するだけでなく，その空間情報を保持したステレオ音声を出力する．これらの研究では，targeted speech が non-target 情報のコンテキストにおいて生成されることを重要視しており，target speech 以外の情報を保持すべきという立場をとっている．我々の研究もこの立場である．

\vspace{-2mm}
\section{Problem definition}
% For the signal $s(t)$, we denote the short-time Fourier transform (STFT) coeeficients as $S(\omega, t)=\textrm{STFT}[s(t)]$ using its uppercase and denote inverse STFT as $s(t)=\textrm{iSTFT}[S(\omega, t)]$.

\subsection{Observation process}
\vspace{-1mm}
% 自然数$N$を話者の人数とする．我々はvoice conversionの対象とする話者の信号$x_1(t)$とその他の$N-1$人の話者の信号$x_2(t),\ldots,x_N(t)$を$N$個のマイクロフォンで観測する．$i\in[N]$番目の音源から$j$番目のマイクロフォンへのインパルス応答を$h_{ij}(t)$とすると，$j$番目のマイクロフォンの観測信号$y_j(t)$は以下のように表される．
% note that 観測信号には$h_{ij}(t)$の畳み込みによって残響や到来方向といった空間情報が付与されている。
% 観測信号を縦に並べたベクトル$[y_1(t),\ldots,y_N(t)]^\top$を$\bm y(t)$と書き，信号$s(t)$のSTFT$\mathrm{STFT}[s(t)]$をその大文字を用いて$S(\omega, t)$と表す．
Let $N$ be a natural number representing the number of speakers.
We observe the speech signal of the target speaker for BSS~(referred to as BSS-target speaker, same as VC-source speaker) $x_1(t)$ and signals from the remaining $N-1$ sources~(non-target signals), denoted as $x_2(t),\ldots,x_N(t)$, using $N$ microphones. 
Let $h_{ij}(t)$ represent the impulse response from the $i$-th sound source to the $j$-th microphone. 
The observed signal $y_j(t)$ at the $j$-th microphone is expressed as follows:
\begin{align}
    y_j(t) = h_{1j}(t) * x_1(t) + \sum_{i=2}^N h_{ij}(t) * x_i(t),
\end{align}
where $*$ denotes convolution.
Note that $y_j(t)$ contains spatial information such as reverberation and direction of arrival~(DoA), imparted by the convolution with $h_{ij}(t)$.
We denote the vector obtained by stacking the observed signals vertically $[y_1(t),\ldots,y_N(t)]^\top$ as $\bm y(t)$ as $\bm y(t)$ and its short-time Fourier transform (STFT) as $\bm Y(\omega, t)$.

\begin{comment}
% voice conversionの対象とする話者を指定する方法は複数考えられるが，本研究では事前にターゲット話者の方向を推定する方法を用いる．
% Among various methods to specify the target speaker for voice conversion, this study employs a method to estimate the target speaker's direction $\theta$ in advance.
\end{comment}
% 本研究では，ターゲット話者の方向へのステアリングベクトル$\bm d(\omega)$は既知とする．これは理想的にはターゲット話者からマイクロホンへの伝達関数であるが，実際には反射や減衰などの要因によって真値から外れた値を取る．
To identify the target speaker, we assume that the steering vector $\bm d(\omega)$ towards the target speaker's direction is known.
Ideally, the vector is equal to the transfer function from the target speaker to the microphones, but in reality, it may differ due to factors such as reflections and attenuation.

\vspace{-1mm}
\subsection{Our purpose}
% 我々の目的は，空間情報や他の話者の音声を保持しつつ1-th話者の話者性を変換した観測信号を生成することである．voice conversionを$\textrm{VC}$とすると$z_j(t)$は以下のように表される：
Our objective is to generate virtual observation signals $z_1(t),\ldots, z_N(t)$ that transform the speaker identity of the BSS-target speaker while preserving spatial information and sounds coming from the other sources. 
Denoting voice conversion as $\textrm{VC}[\cdot]$, $z_j(t)$ for each $j=1,\ldots,N$ is expressed as follows:
\begin{align}
\label{eq:ideal_remixing}
    z_j(t) = h_{1j}(t) * \textrm{VC}[x_1(t)] + \sum_{i=2}^N h_{ij}(t) * x_i(t).
\end{align}
Through $\textrm{VC}[\cdot]$, $x_1(t)$ is transformed into the VC-target speaker.

\vspace{-2mm}
\section{Proposed method}
\label{sec:proposed_method}

\begin{figure}[t]
    \centering
    \includegraphics[width=1.0\linewidth]{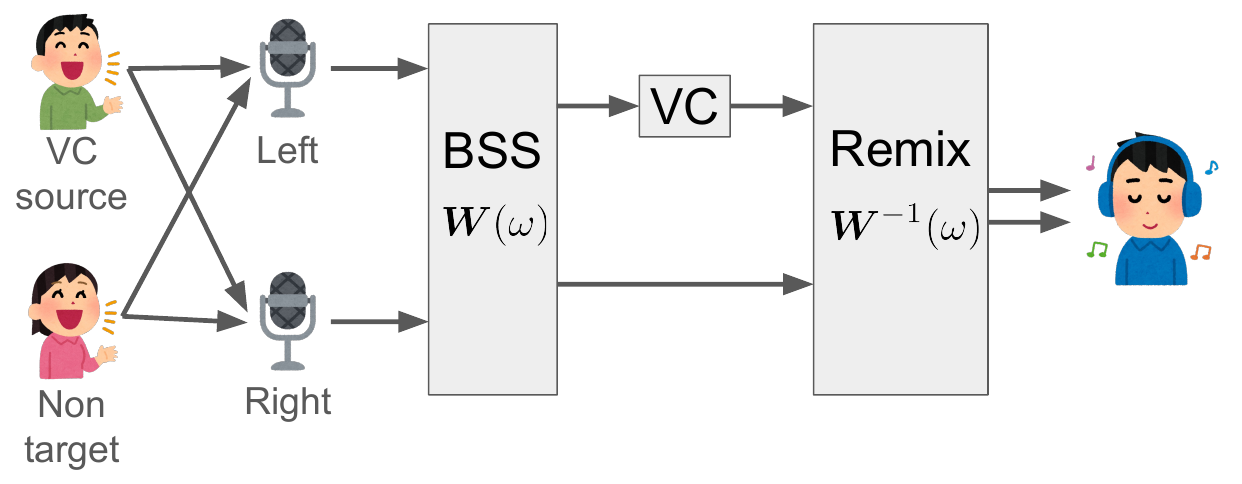}
    \vspace{-6.5mm}
    \caption{Proposed method procedure: We record mixed audio from multiple speakers using multiple microphones, apply BSS to extract the voice of the BSS-target speaker (the same as the VC-source speaker), apply VC exclusively to this voice, and then output the remixed multi-channel signal.}
    \label{fig:procedure}
    \vspace{-3mm}
\end{figure}

% 我々の提案手法の流れは図に示す通り，speech separation, voice conversion~(VC), remixing の過程からなる．
% As shown in Fig.~\ref{fig:procedure}, our proposed method consists of the following processes: BSS, VC, and remixing.

A straightforward approach is to apply the VC model to each channel individually. 
However, applying VC to each channel separately results in inconsistent VC outputs across channels, failing to maintain the coherence of the multi-channel signal. 
Therefore, our proposed method applies VC to the monaural speech signal of the VC-source speaker.
It consists of the processes of BSS, VC, and remixing, as shown in Fig.~\ref{fig:procedure}.

\vspace{-1mm}
\subsection{Blind source separation}
\label{sec:method:speech_separation}
\vspace{-1mm}

% 本研究で用いる voice conversion モデルは単一話者の音声を入力として変換後の音声を出力するモデルであるため，複数の信号が混合している観測信号から所望の話者の音声のみを抽出する必要がある．
% フロントエンドの音声強調として、BSSを用いてVC-source speakerのモノラル信号音声を取得する。
% 我々はBSS-target speakerの声を強調するのと同時にnoiseの声も抽出したいため、BSSを適用して混合音声を分離する。
%%%% 本研究ではマルチチャネル音源分離によって混合信号の分離を実施する．
% 既存のBSS手法は残響やモデル化誤差に頑健な分離を達成できるため，本研究ではこれを用いて混合信号の分離を行う．
% ただし完全にblindな分離では出力順序を特定することができないため，空間正則化を用いたBSS such as GC-IVAによって1番目のチャネルにvoice conversionのターゲットの話者の音声が出力されるようにする．
\begin{comment}
BSSは残響や散乱体によって空間情報のモデル化が困難な場合にもロバストに音声強調を達成するため、本研究ではこれを用いて混合信号の分離を行う。
ただし完全にblindな分離では出力順序を特定することができないため，空間正則化を用いたBSS such as GC-IVAによって1番目のチャネルにvoice conversionのターゲットの話者の音声が出力されるようにする．
\end{comment}
We apply BSS to separate the mixed audio because we want to enhance the voice of the BSS-target speaker while also extracting the non-target signals.
However, since completely blind separation cannot identify which signal corresponds to the BSS-target speaker, we utilize spatially regularized BSS (as mentioned in Section~\ref{sec:related_work}) to ensure that the voice of the target speaker for VC is output in the first channel.

The separated signals obtained with BSS are represented as $\bm V(\omega)\bm Y(\omega, t)$, where $\bm V(\omega)$ is an $N \times N$ separation matrix.
Each component of this introduces a constant scaling factor for each frequency, resulting in signal distortion.
To mitigate this distortion, we employ a technique known as projection back~(PB)~\cite{murata2001approach} and solve the following optimization problem:
\begin{align}
& \underset{\bm W(\omega)} {\text{minimize}} &&  
  \mean{\left\lVert \bm 1_{N}^\top\bm W(\omega)\bm Y(\omega, t) - Y_1(\omega, t) \right\lVert^2}, \\
& {\text{subject to}} && \exists k_1(\omega),\ldots,k_N(\omega), \notag \\
&                     && \bm W(\omega) = \textrm{diag}\{k_1(\omega),\ldots,k_N(\omega)\} \bm V(\omega)
\end{align}
where $Y_1(\omega, t)$ denotes the first component of $\bm Y(\omega, t)$ and $\bm 1_{N}$ represents an $N$-dimensional all-ones column vector.
Finally, with a separation matrix $\bm W(\omega)$, we obtain the separation result as $\hat {\bm X}(\omega, t)=\bm W(\omega)\bm Y(\omega, t)$.

Please note that while PB mitigates distortion, it does not completely resolve the issue of scaling factors introduced during separation. 
As a result, the $i$-th component $\hat{X}_i(\omega, t)$ corresponds to the observed signal $h_{11}(t) * x_1(t)$, rather than the original source signal $x_1(t)$.
This issue becomes significant in the remixing process.

\vspace{-1.5mm}
\subsection{Voice conversion}
\vspace{-0.5mm}
% ターゲット話者の音声にvoice conversionを適用する．
% related workで述べたようにノイズや残響を保つvoice conversionも提案されているが，本研究ではcleanなspeechを出力するvoice conversion modelを利用する．cleanな音声を出力する問題設定は広く研究されており利用可能なコード・学習済みモデルが多いため，これらのリソースを活用することによるメリットを享受できる．
\begin{comment}
Among voice conversion methods preserving noise and reverberation have been investigated, this study uses voice conversion method generating clean speech due to the widesperead reseach and the availability of numeraous codes and pretrained models.
By leveraging these resources, it becomes possible to extend the application beyond speaker conversion to other problem settings, such as emotion conversion.
\end{comment}
% ここではターゲット話者の音声にvoice conversionを適用した$\textrm{VC}[\hat X_1(\omega, t)$を出力する．
% We denote STFT of the output as $\textrm{VC}[\hat X_1(\omega, t)]$.
% VC-target speakerに変換された音声のSTFT$\textrm{VC}[\hat X_1(\omega, t)$とする．
% VCによって$hat X_1(\omega,t)$をVC-target speakerに変換し，we denote the 出力 as $(ここに式)$
We apply VC to the extracted speech signal \(\hat{X}_1(\omega,t)\), which corresponds to the VC-source speaker.
Although VC methods that preserve noise and reverberation have been proposed, this study employs a VC model focused on producing clean monaural speech.
The task of generating clean monaural audio has received widespread attention, resulting in a variety of available codes and pretrained models. 
Transforming $\hat X_1(\omega, t)$ to the VC-target speaker using VC, we denote the STFT of the output as $\textrm{VC}[\hat X_1(\omega, t)]$.

\subsection{Remixing}
\label{sec:method:remix}
\begin{comment}
% 我々の利用できる空間情報は分離行列の逆行列$\bm W(\omega)^{-1}$であり，この$(i,j)$成分は$i$番目の音源について$1$-thマイクロホンから$j$-thマイクロホンへの相対伝達関数を表す．
% これは$1$-thマイクロホンにおける残響の除去をimplicitに行うことになり，clean 信号であり残響を含まない$\textrm{VC}[\hat X_1(\omega, t)]$に適用すると音声品質が劣化する．
The spatial information available to us is the inverse matrix of the separation matrix, denoted as $\bm W(\omega)^{-1}$.
Its $(i,j)$ element indicates the relative transfer function from the $1$-th microphone to the $j$-th microphone for the $i$-th source.
This implicitly performs dereverberation at the $1$-th microphone, and degrade the speech quality when applied to $\textrm{VC}[\hat X_1(\omega, t)]$, which is clean signal without reverberation.
Therefore, it is necessary to estimate the transfer function $H_{11}(\omega)$ from the target speaker to the $1$-th microphone in order to reproduce the reverberation.
\end{comment}
% voice conversionの出力$\textrm{VC}[\hat X_1(\omega, t)]$と分離信号$\hat{\bm X_{2:N}}(\omega,t):=[\hat X_2(\omega,t),\ldots,\hat X_N(\omega,t)]^\top$に空間情報を付与し，所望のマルチチャネル信号をobtainする．
We impart spatial information to VC output $\textrm{VC}[\hat X_1(\omega, t)]$ and non-target signals $\hat{\bm X}_{2:N}(\omega):=[\hat X_2(\omega,t),\ldots,\hat X_N(\omega,t)]^\top$, obtaining desired multi-channel signal.
Since remixing is the reverse process of separation, a straightforward approach is to apply the inverse matrix of the separation matrix, denoted as $\hat {\bm A}(\omega):=\bm W(\omega)^{-1}$. 
In fact, the $(i, j)$ element $\hat A_{ij}(\omega)$ represents the relative transfer function from the $1$-th microphone to the $j$-th microphone for the $i$-th sound source, so by applying it to $\hat{\bm X}_{2:N}(\omega)$, spatial information can be reproduced.
This approach output the following signal:
\begin{align}
\label{eq:inverse}
    \hat{\bm Z}(\omega,t) = \hat {\bm A}(\omega)
    \begin{bmatrix}
     \textrm{VC}[\hat X_1(\omega, t)]
     \\
     \hat{\bm X}_{2:N}(\omega,t)
    \end{bmatrix}.
\end{align}

However, this method is inappropriate for $\textrm{VC}[\hat X_1(\omega, t)]$. 
The relative transfer function includes the inverse filter of the transfer process from the speaker to the $1$-th microphone. 
Unlike $\hat{\bm X}_{2:N}(\omega)$, which is affected by the transfer function $h_{i,1}(t)$, $\textrm{VC}[\hat X_1(\omega, t)]$ is clean signal.
Applying $\hat {\bm A}(\omega)$ introduce significant distortion due to the instability of the inverse filter.
Therefore, for $\textrm{VC}[\hat X_1(\omega, t)]$, an alternative method is required to reproduce the spatial information.

Another method to explicitly reproduce the DoA using the steering vector $\bm d(\omega)$ can be considered. 
The steering vector $\bm d(\omega)$ models the direct path and calculates the transfer function from the VC-source speaker to the $i$-th microphone. 
Although this method cannot reproduce the reverberation, it is expected that reproducing the DoA will sufficiently capture the essential spatial information.
Denoting the $i$-th column of $\hat{\bm A}(\omega)$ as $\hat{\bm a}_i(\omega)$ and defining $\hat {\bm A}_{2:N}:=[\hat{\bm a}_2(\omega),\ldots,\hat{\bm a}_N(\omega)]$, the output signal can be expressed by the following equation:
\begin{align}
\label{eq:steering}
    \hat{\bm Z}(\omega,t) = 
    \begin{bmatrix}
        \bm d(\omega) & \hat {\bm A}_{2:N}(\omega)
    \end{bmatrix}
    \begin{bmatrix}
     \textrm{VC}[\hat X_1(\omega, t)]
     \\
     \hat{\bm X}_{2:N}(\omega,t)
    \end{bmatrix}.
\end{align}

\section{Discussion}
\label{sec:method:discuss}
% 本タスクにおいて生じる問題を考察する．まず，voice conversion への入力は，残響や分離誤差として含まれるnon-taget speakerの声といった要因で劣化している．昨今のVCは実環境でも動くロバスト性があるためこの劣化は大きな問題にならないと考えられるが，non-target sound sourceが多い環境や残響時間の長い環境など劣化が激しくなるとノイズロバストなVCが必要になると考えられる．
% We consider the problems that arise in this task.
% This section lists the problems to be solved in this task.

% this sectionは、Section~\ref{sec:proposed_method}で提案したbaselineにおける問題をlistする
This section lists the problems in the our proposed method.

\begin{comment}
Firstly, in voice conversion, the input is degraded by factors such as reverberation and non-target speaker's voice included as separation error. 
Recent advances in voice conversion exhibit robustness even in real-world environments, so this degradation is not considered a significant issue.
However, in environments with many non-target sound sources or long reverberation times, degradation becomes more severe, and noise-robust voice conversion is deemed necessary.
\end{comment}
\textbf{Robustness of voice conversion.} 
The input to the VC is degraded due to being the distant speaker's voice and the noise introduced by separation errors. 
Therefore, the VC model needs to be robust. 
Since our method utilizes a monaural VC model, we can address this issue by benefiting from the robustness of modern VC models~\cite{huang2022toward, igarashi2024noiserobust} designed to operate effectively in real-world scenarios with monaural audio.

% However, unlike monaural audio, multi-channel audio can involve numerous non-target sound sources and long reverberation times, requiring a level of robustness different from traditional voice conversion.

% 本研究ではターゲット話者を指定するのにアレイと話者との位置関係を仮定したが，他にもspeaker identificationなどsource signalのみを手がかりに特定する方法も考えられる．実験で示す通りステアリングベクトル$\bm d(\omega)$の推定なしに再合成を行うことは困難であるため，この場合はmusicなどの音源定位法を組み合わせて$\bm d(\omega)$の推定を行うことが必要と考えられる．
\begin{comment}
In this study, we assumed an array and the spatial relationship between the array and the speaker to specify the target speaker, but alternative methods such as speaker identification based solely on source signals are also conceivable.
As shown in the experiments, it is difficult to perform remixing without estimating the steering vector $\bm d(\omega)$, so in such cases, it is necessary to combine methods such as source localization techniques used in music to estimate $\bm d(\omega)$.

Spatial VCの入力には複数の信号が含まれるため、そのうちどれがVC-source speakerであるかを特定する必要がある。
我々の手法ではthe spatial relationship between the microphones and the speaker を仮定することでこれを解決している。
another approachはspeaker identificationによって特定する方法であり、この場合空間情報を用いずにtrackingを実現できる。
\end{comment}
\textbf{Target speaker tracking.} 
Since the input to spatial VC contains multiple signals, it is necessary to identify which one is the VC-source speaker. 
Our method addresses this by assuming the spatial relationship between the microphones and the speaker. 
Another approach is to use speaker identification, which allows tracking without relying on spatial information.

% 分離が失敗した場合の影響をdiscussする．non-targtet speaker同士がうまく分離されないことは起こりうるが，これについては最終的にはremixにおいて逆行列をかけることで吸収されるため影響はない．BSS-target speakerの音声が他の信号に混じることがあり，これは最終的な信号にも混入することとなる．これはvoice conversion出力と同じタイミングで同じ発話内容を話すため，隠れて知覚されないと考えられる．
% We will now discuss the impact of separation failure. It is possible for non-target speakers to not be separated effectively, but this issue will ultimately be absorbed when applying the inverse matrix during remixing, rendering it inconsequential. There may be cases where the speech of the BSS-target speaker mixes with other signals, which can then contaminate the final output. However, since this occurs when speaking the same content at the same timing as the voice conversion output, it is likely to go unnoticed and not perceptible.
% 音源分離の分離誤差で信号が混入することの影響を議論する。
% $\hat X_1(\omega, t)$ に他のノイズが混入する影響は、上記で議論した通り、ノイズロバストなVCを用いることで解決できる。
% non-target signals 間の分離誤差については、後段のremixingで吸収されるので全く影響がない。
% non-target signals の出力である$\hat{\bm X}_{2:N}(\omega,t)$にVC-source speakerの声が混入する場合、最終出力にVC-source speakerとVC-target speakerの両方の音声が含まれることになる。
% この2つの音声は同じ内容を同時に発話するため、混入する成分が十分小さければマスクされて知覚されないと考えられる。

\textbf{Impact on signal contamination.}
Let us consider a case that the voices of each speaker are not sufficiently separated and the voice of one speaker slightly mixes with another's.
The effect of non-target signals contaminating $\hat X_1(\omega, t)$ can be addressed by using noise-robust VC, as discussed above. 
The separation error between non-target signals has ultimately no impact because it is absorbed in the subsequent remixing stage. 
If the voice of the VC-source speaker is mixed into the output of the non-target signals $\hat{\bm X}_{2:N}(\omega, t)$, the final output will contain both the VC-source speaker and the VC-target speaker. 
Since these two voices are speaking the same content simultaneously, it is expected that if the contaminating component is sufficiently small, it will be masked and not perceived.

\begin{comment}    
% 次にremixについて議論する．ストレートに考えれば分離行列の逆行列を用いることで混合過程が再現できると考えられるが，実際には4.3.で議論した要因によって自然性が低下する as 実験で示す通り．これはspatial voice conversionが単なる音源分離とvoice conversionの結合タスクではなく音響伝達関数の推定を必要としていることを示している．
\textbf{Preserving spatial information via remixing.} 
% While it may seem straightforward to reproduce the mixing process using the inverse matrix of the separation matrix, 
While it may seem a straightforward approach of remixing involves reversing the separation process,
in practice, this may decrease the naturalness as discussed in Section 4.3. This indicates that spatial VC requires more than just a concatenation of BSS and VC tasks; it necessitates the estimation of acoustic transfer functions.

Section~\ref{sec:method:remix}で議論した通り，分離行列を用いるremixing方法(Eq.~\ref{eq:inverse})は品質劣化を招く．
これはPBの説明で議論したBSSの定数倍の自由度によるものであり、具体的なBSS手法に依存せず本質的な課題である。

これはSection~\label{sec:method:speech_separation}で説明したconstant scaling factorに起因し、具体的なBSS手法に依存しない本質的な課題である。

一方でsteering vectorを用いる方法(Eq.~\ref{eq:steering})は実環境において残響や散乱体の影響でモデル化誤差が大きくなり，空間情報を適切に捉えることができない．
品質を劣化させずに空間情報をreproduceすることは困難な課題であり，そのことを実験で確認する．
\end{comment}
\textbf{Balancing audio quality and spatial information in remixing.}
As discussed in Section~\ref{sec:method:remix}, the remixing approach using the separation matrix (Eq.~\ref{eq:inverse}) leads to quality degradation. 
This issue arises from the constant scaling factor of BSS explained in Section~\ref{sec:method:speech_separation} and is an inherent challenge regardless of the specific BSS method used.
On the other hand, in practical settings, the method utilizing the steering vector (Eq.~\ref{eq:steering}) faces challenges in capturing spatial information accurately due to modeling errors caused by reverberation and scattering.
Therefore, reproducing spatial information without degrading quality proves to be a challenging task.

\section{Experimental evaluations}
We conducted experimental evaluations to identify challenges in achieving spatial VC.

\subsection{Data and settings}
% シミュレーションデータに対する実験を通して，提案手法がspatial voice conversionを実現することとその上での課題について確認する．
% We conducted experimantal evaluations on simulation data to identify the challenges in spatial voice conversion.
% Additionaly, we compared multiple remixing methods to address challenges associated with this task.

\begin{figure}[t]
    \centering
    \includegraphics[width=0.8\linewidth]{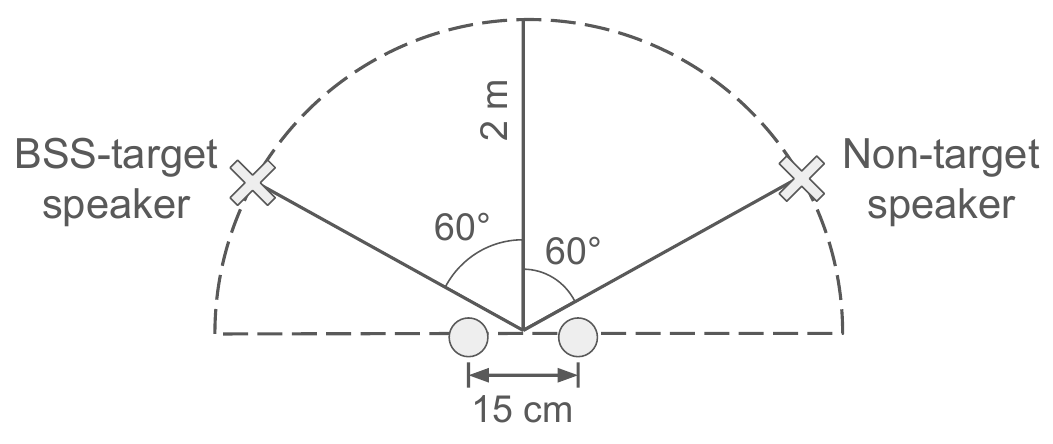}
    \vspace{-3mm}
    \caption{Configurations of speakers and microphones, where ``×'' and ``o'' denote speakers and microphones, respectively.}
    \label{fig:experimental_setup}
    \vspace{-4mm}
\end{figure}

% シミュレーションデータの作成には JVS コーパスのnonparalell 30のsubsetからJVS001～JSS010の10 speakers(male 5, female 5)を利用した．
% 観測信号はpyroomacousticsを用いて2話者の音声を2チャンネルで収録することをシミュレーションして作成する．
% 音声は約3割がオーバーラップするように時間をずらした，実際の会話では発話は完全にオーバーラップしないため．
% 話者はvoice conversionのソースとなる話者，妨害話者の２人で，それぞれ$x_1(t),x_2(t)$に対応し，異なる性別の話者を選んだ．
% マイクロホンとソース話者・妨害話者の位置関係は図に示す．マイクロホンの間隔は$15cm$とする．部屋のサイズは9m x 7.5m x 3.5mであり，reflection coefficient は$0.5$に設定し，reverberation time($RT_{60}$)は約$200ms$であった．
To generate simulation data, we utilized speech data from a nonparallel30 subset of the JVS corpus~\cite{JVS}, specifically from jvs001 to jvs010, 10 speakers (5 males and 5 females).
The observed signals were generated by simulating two-channel recordings of two speakers using pyroomacoustics~\cite{scheibler2018pyroomacoustics}, representing the left and right channels.
The speakers consist of the BSS-target and non-target speakers, corresponding to $x_1(t)$ and $x_2(t)$, respectively. 
They were chosen to be speakers of different genders, allowing listeners to identify the target speaker more effectively.
Utterance durations of the two speakers were overlapped by $50\%$, reflecting the infrequent occurrence of complete overlap in real conversations.
The positions of the speakers and microphones are shown in Fig.~\ref{fig:experimental_setup}.
The interval of the microphones was set at $15~\si{cm}$.
The room size is $9~\si{m}\times 7.5~\si{m}\times 3.5~\si{m}$.
The reflection coefficient $r$ was set to $0.5$, and the reverberation time ($RT_{60}$) was approximately $200~\si{ms}$.

% proposed methodで用いる$\bm d(\omega)$はマイクロホンアレイの中心からターゲット話者への角度$\theta$を用いてfar-field仮定で計算されたステアリングベクトルを利用する．
% BSS には GC-IVA~\cite{li2020geometrically} を利用し，ステアリングベクトル$\bm d(\omega)$によるヌル制約を重み$\lambda=1$で加算した評価関数をauxiliary function-based method~\cite{ono2011stable}によって最小化する．
% BSS にはGC-IVA~\cite{li2020geometrically} を利用し，評価関数にヌル制約usingステアリングベクトル$\bm d(\omega)$を可変重み$\lambda$で加えた．最小化はauxiliary function-based method~\cite{ono2011stable}によって行い，$\lambda=1$として50反復行なった後$\lambda=0$として50反復した．
% voice conversion model はpublicli availableなコードであるDDSP-SVCのver 5.0を利用し，JVSコーパスのparallel 100subsetで$100k$ステップ学習した．
% voice conversionのターゲットスピーカーは，ターゲット話者と同じ性別の話者をjVS001～JVS010からランダムに選択した．
% projection backのターゲットのchannelとしてはleftを選択した．
All the speech signals were sampled at $24~\si{kHz}$. The STFT was computed using a Hann window whose length was set at $4096$ samples, and the window shift was $2048$ samples.
The steering vector $\bm d(\omega)$ used in our proposed method was calculated under the far-field assumption based on the angle $\theta$ from the center of the microphone array to the target speaker’s position
For BSS, we employed GC-IVA~\cite{li2020geometrically}, augmenting the evaluation function with a null constraint using the steering vector $\bm d(\omega)$ with a variable weight $\lambda$.
The minimization process was carried out using the auxiliary function-based method~\cite{ono2011stable}, iterating $50$ times with $\lambda=1$, followed by $50$ iterations with $\lambda=0$.
The VC model was implemented using the publicly available code of DDSP-SVC~\cite{DDSP-SVC} version 5.0, and it was trained for $100$k steps on the parallel100 subset from the JVS corpus~\cite{JVS}.
The VC-target speakers were randomly selected from jvs001 to jvs010, ensuring they are speakers of the same gender as the VC-source speakers.
We chose the left channel as the PB-target channel.

% 事前実験では，式5を用いたremixingが著しく低い性能であった．このことを確認するため，以下の3つのremixing手法を評価する．
% inverse-remixing: 提案手法によってsatial voice conversionを行い，remixingは式~\ref{eq:inverse}で示したように分離の逆過程として実施する．BSSを用いた推定によって残響などの要因を考慮できることが期待できるが，Section~\ref{sec:method:remix}で述べた逆フィルタの問題から自然性が損なわれると考えられる．
% steering-remixing: 提案手法によってspatial voice conversionを行い，remixingは式~\ref{eq:steering}で示したようにステアリングベクトルを用いて実施する．実際の場面では残響や散乱対の影響から空間情報の再現性が落ちるものの，自然性を高く保つと考えられる．
% ideal: 比較対象として，式2で示した所望の音声をobtainした．VCは分離出力ではなく理想的な収録音声を受け取り，空間情報はシミュレーションによって付加された．
% 以上の方法で作成した音声に含まれるターゲット話者の音声について，自然性と，正解との類似性について主観評価実験を実施した．
% 自然性の評価は4つの手法で作成した音声サンプルのステレオ音声，monoral 音声( Left, Right channel) について実施した．100 listnerが参加し，各listnerは20サンプルを5段階で評価した．ただしrange equlization bias~\cite{cooper23_interspeech}を避けるため，各lisnterは評価前にdummyとして6サンプルを評価し，範囲の決定を行った．
% また，idealとそれ以外の3手法のステレオ音声，モノラル音声（left, right channel) について類似度の主観評価を実施した．100 listnerが参加し，各listnerは4つのdummyペアを聞いたのちに10ペアの類似度を5段階で評価した．
% In preliminary experiments, it was suggested that remixing is challenging for this task. 
% To evaluate this difficulty, we compared the following remixing methods:
%%%%%%%%%%%%%%%%%%%%%%%%%%%%%%%%%%%%%%%%%%%%%
% Section~\ref{sec:method:discuss}で議論したremixingの問題を確認するため，以下の3つの方法を評価する．
To verify the issues discussed in Section~\ref{sec:method:discuss}, we evaluated the following three methods:
\begin{itemize}
    \item ``\inverse'': 
    We performed spatial voice conversion using the proposed method, and the remixing was carried out as the inverse process of separation, as outlined in Eq.~\ref{eq:inverse}. 
    Utilizing BSS for estimating spatial information allows for the consideration of factors such as DoA and reverberation.
    However, this method may compromise acoustic quality due to issues with the inverse filters, as discussed in Section~\ref{sec:method:remix}.
    \item ``\steering'': 
    We performed spatial voice conversion using the proposed method, and remixing was conducted using the steering vector, as outlined in Eq.~\ref{eq:steering}.
    This method ignores reverberation, a significant factor in real-world scenarios, so the reproducibility of spatial information may decrease. 
    However, it is expected to maintain audio quality, unlike ``\inverse''.
    \item ``\ideal'': 
    As a reference for comparison, we obtained the desired audio as indicated in Eq.~\ref{eq:ideal_remixing}. 
    In this method, VC takes an ideal recorded speech instead of BSS output, and spatial information is added through simulation.
\end{itemize}
\begin{comment}    
For the speech of the VC-target speaker included in the generated audio using the above methods, subjective evaluation experiments were conducted in two metrics.
Firstly, naturalness evaluation involved stereo and monaural (left, right channel) samples created with the three methods. $100$ listeners participated, each rating 20 samples on a 5-point scale. 
To avoid range equalization bias~\cite{cooper23_interspeech}, each listener evaluated six dummy samples before the actual evaluation to establish the rating scale.
% こっからしたはあれ
% Log-Determinant Divergence (LDD) kulis2006learning
Additionally, we evaluated subjective similarity between the ``\ideal'' and the other three methods. 
Evaluations were conducted on stereo and monaural (left, right channel) audio and $100$ listeners participated, rating the similarity of 10 pairs on a 5-point scale after listening to four dummy pairs.
We calculated the mean opinion score (MOS) for each evaluation.
\end{comment}
\begin{comment}
出力音声の品質及び空間情報の再現度をevaluateするため，自然性の主観評価と空間相関行列の客観評価を行なった．
100 listener が VC-target speaker の音声の自然性を5段階で評価し，calculated mean opinion score~(MOS).
各聴取者は3つの手法で作成した20サンプルを評価し，その前にdummyとして6サンプルを評価することでrange equlization bias~\cite{cooper23_interspeech}を回避した．
ステレオ音声の評価に加えてモノラル (left, right channel) の評価を行なった，which チャネルが品質低下に繋がってるか調べるために．
客観評価として，``\ideal``から``\inverse''及び``\steering''へのSCMのLog-Determinant Divergence (LDD)~\ref{kulis2006learningw}を計算した．
はずれ値の影響を除去するために，第一四分位数〜第三四分位数を抽出して平均・標準偏差（SD）を計算した．
\end{comment}
To assess the quality of the output audio and the accuracy of spatial information reproduction, we conducted subjective evaluations for naturalness and objective evaluations using the spatial correlation matrix (SCM).
A total of 100 listeners rated the naturalness of the VC-target speaker's voice on a five-point scale.
Each listener assessed $20$ samples, preceded by $6$ dummy samples to calibrate range equalization bias~\cite{cooper23_interspeech}.
Then, we calculated mean opinion score (MOS) for each method. 
In addition to evaluating stereo audio, assessments for monaural channels (left and right) were conducted to investigate which channel contributes to quality degradation. 
For objective evaluation, we computed the log-determinant divergence (LDD)~\cite{kulis2009low} of SCM from ``\ideal'' to both ``\inverse'' and ``\steering''  as LDD of SCM serves as a metric to represent the estimation error in spatial information~\cite{kubo2020blind}.
To mitigate the impact of outliers, we extracted data from the first quartile to the third quartile and calculated the mean and standard deviation (SD).

\subsection{Results}
\subsubsection{Acoustic quality}

\begin{table}[t]
% 各手法でsynthesizeした音声の自然性MOSと標準偏差．
\caption{
    MOS on naturalness of VC-target speaker's speech with each method and SD of MOS.
    A higher value indicates better audio quality~($\uparrow$).
}
\vspace{-2mm}
\label{table:naturalness}
\centering \footnotesize
\begin{tabular}{c||c|c|c}
% Methods & Stereo & Left~(PB target) & Right \\ \midrule
\multirow{2}{*}{Methods} & \multirow{2}{*}{Stereo} & Monaural-Left & \multirow{2}{*}{Monaural-Right} \\ 
&  & (PB target) &  \\ \midrule
\ideal &  $ 4.254 \pm 0.060 $ &  $ 4.062 \pm 0.060 $ &  $ 4.102 \pm 0.058 $ \\ \midrule
\inverse &  $ 1.992 \pm 0.064 $ &  $ 4.142 \pm 0.060 $ &  $ 1.388 \pm 0.042 $ \\
\steering &  $ 3.405 \pm 0.062 $ &  $ 3.975 \pm 0.066 $ &  $ 3.132 \pm 0.070 $ 
\end{tabular}
\vspace{-5mm}
\end{table}

Table~\ref{table:naturalness} presents MOSs on naturalness.
In the stereo results, ``\inverse'' exhibited a significantly lower score compared to ``\steering,'' indicating that the remixing process in ``\inverse'' caused a noticeable degradation in audio quality.

Comparing the monaural-left results, which are the target channel for PB, ``\inverse'' achieved naturalness comparable to the other methods.
Achieving a high score comparable to ``\ideal'' suggests that the degradation of VC input does not negatively impact the quality of VC output. 
This is due to the robustness of VC, as discussed in Section~\ref{sec:method:discuss}.
In contrast, in the monaural-right channel, ``\inverse'' yielded notably lower results.
This outcome highlights that the relative transfer functions in this method degrade audio quality significantly as discussed in Section~\ref{sec:method:remix}.
Therefore, from an audio quality perspective, ``\steering'' should be employed.

\vspace{-1.4mm}
\subsubsection{Spatial information}
\vspace{-0.6mm}

\begin{comment}
\begin{table}[t]
% 各手法でsynthesizeした音声の自然性MOSと標準偏差．
\caption{
    LDD.
}
\label{table:ldd}
\centering \footnotesize
\begin{tabular}{c||c||c|c|c}
Methods & Mean & 1st quartile & 2nd quartile & 3rd quartile \\ \midrule
\inverse &  $ 1.686 \pm 3.298 $ &  $ 0.774 $ &  $ 0.969 $ & $1.223$\\
\steering &  $ 3.572 \pm 2.410 $ &  $ 2.261 $ &  $ 2.988 $  & $3.974$
\end{tabular}
\vspace{-2mm}
\end{table}
\end{comment}
\begin{table}[t]
% ``\ideal'' から各手法へのSCMのLDDとそのSD．低いほど空間情報を正確に再現できていることを表す．
\caption{
    LDD of SCM from ``\ideal'' to each method and its SD in each reflection coefficient $r$.
    A lower value indicates a more accurate reproduction of spatial information~($\downarrow$).
}
\vspace{-2mm}
\label{table:ldd}
\centering \footnotesize
\begin{tabular}{c||c|c|c}
Methods & $r=0.2$ & $r=0.5$ & $r=0.8$  \\ \midrule
\inverse &  $0.468 \pm 0.074$ & $ 0.979 \pm 0.147$ &  $2.058 \pm 0.270$ \\
\steering & $1.990 \pm 0.367$ & $ 3.097 \pm 0.552$ &  $4.666 \pm 0.535$
\end{tabular}
\vspace{-3mm}
\end{table}

\begin{comment}    
Table~\ref{table:ldd}にLDDを示す．Naturalness 評価と同じ条件($r=0.5$)に加え，異なるreflection coefficient $r$ の設定でも評価を行なった．
$r=0.2, 0.8$はそれぞれ$RT_{60}$が約$70~\si{ms},400~\si{ms}$に対応する．

異なる残響時間での比較を行うと，いずれの手法も残響時間が伸びるにつれて性能が低下していることが分かる．
これは残響時間が伸びるにつれBSSが困難になり，BSSで推定された空間情報の精度が低下することに起因すると考えられる．
さらに，いずれの残響時間においても``\inverse''は``\steering''よりも優れており，残響時間が伸びるほどこの差は広がっている．
この違いは``\inverse''が空間情報を柔軟に推定しているのに対し``\steering''は事前決定した限られたモデルを用いていることに起因すると考えられる．
すなわち，``\steering''は考慮外の要因が加わることで空間情報の推定精度が低下する手法であると言える．
これは空間情報を柔軟に推定する``\inverse''と異なり，事前に定めたモデルを用いてtarget  speakerの空間情報を構成する``\steering''では考慮外の要因が増大することで空間情報再構成の性能が低下することを示している．
practical setting では散乱体の影響などが加わり$\bm d(\omega)$のモデル化誤差が増長するため， ``\steering''の空間情報の再現精度はさらに落ちることが予想される．
以上の結果から空間情報の結果では ``\inverse'' を用いるべきと言えるが，これは audio quality の観点での結果と反対である．
\end{comment}
Table~\ref{table:ldd} presents the LDD values.
In addition to the condition used for naturalness evaluation ($r=0.5$), we conducted assessments with different reflection coefficient ($r$). 
Specifically, $r=0.2$ and $r=0.8$ correspond to $RT_{60}$ values of approximately $70~\si{ms}$ and $400~\si{ms}$, respectively. 

Comparing different reverberation times, the LDD of both methods deteriorated as the reverberation time increased.
This decline is attributed to the growing challenge of BSS with longer reverberation times, resulting in a decrease in the accuracy of spatial information estimated by BSS.

Furthermore, across all reverberation time settings, ``\inverse'' outperformed ``\steering,'' and this difference widens with the longer reverberation time.
This difference is attributed to the flexibility in spatial information estimation of ``\inverse,'' while ``\steering'' relies on a predetermined and limited model.
In other words, ``\steering'' is a method where the estimation accuracy of spatial information decreases when external factors come into play, indicating that it is sensitive to unaccounted factors.
In practical settings, where reverberation is an important factor for spatial information, ``\steering'' may particularly struggle with spatial information reproduction, amplifying errors in the steering vector.
Therefore, the results suggest that ``\inverse'' should be favored from a spatial information perspective. 
However, this contradicts the conclusion from the audio quality perspective.

\section{Conculusion}
We proposed a novel task called spatial voice conversion. 
This task extends voice conversion (VC) from single-channel signals to multi-channel signals, aiming to perform VC while preserving spatial information and non-target signals. 
In this paper, we integrate existing blind source separation (BSS) methods with VC models to address this task, highlighting the challenges posed by remixing. 
In our discussion, we organized the challenges of this task. 
Although the experimental conditions were limited, we found that balancing audio quality and spatial information reproduction presents a fundamental difficulty. 
Therefore, a crucial question arises for future work: \textit{Is it possible to achieve spatial voice conversion that balances both audio quality and spatial information reproduction?}

\section{Acknowledgements}
This work is supported by Research Grant S of the Tateishi Science and Technology Foundation.

% \newpage
\bibliographystyle{IEEEtran}
\bibliography{bib/bss,bib/vc}

\end{document}